\documentclass[conference]{IEEEtran}
\usepackage{cite}
\usepackage{amsmath,amssymb,amsfonts}
\usepackage{algorithmic}
\usepackage{graphicx}
\usepackage{textcomp}
\usepackage{amsmath,amssymb,amsfonts}
\usepackage{algorithmic}
\usepackage{graphicx}
\usepackage{textcomp}
\usepackage{multicol}
\usepackage{hyperref}
\usepackage{multirow}
\usepackage{booktabs}
\usepackage{tabularray}
\usepackage{stfloats}
\usepackage{pdflscape}
\usepackage{tabularx}
\usepackage[table,x11names]{xcolor}
\usepackage{adjustbox}
\usepackage{rotating}
\usepackage{svg}
\usepackage{hhline}
\usepackage{balance}
\usepackage{csquotes}
\usepackage[T1]{fontenc}
\usepackage{csquotes}
\usepackage[many]{tcolorbox}    	

\def\BibTeX{{\rm B\kern-.05em{\sc i\kern-.025em b}\kern-.08em
    T\kern-.1667em\lower.7ex\hbox{E}\kern-.125emX}}
\begin{document}

\title{Tracing the Lifecycle of Architecture Technical Debt in Software Systems: A Dependency Approach}


\author{
    \IEEEauthorblockN{
        Edi Sutoyo\IEEEauthorrefmark{1}\IEEEauthorrefmark{2}, Paris Avgeriou\IEEEauthorrefmark{1}, Andrea Capiluppi\IEEEauthorrefmark{1}
    }
    \IEEEauthorblockA{\IEEEauthorrefmark{1}Bernoulli Institute, University of Groningen, Groningen, The Netherlands}
    \IEEEauthorblockA{\IEEEauthorrefmark{2}Department of Information Systems, Telkom University, Bandung, Indonesia}
    Email: \{e.sutoyo, p.avgeriou, a.capiluppi\}@rug.nl
}

\maketitle

\begin{abstract}
Architectural technical debt (ATD) represents trade-offs in software architecture that accelerate initial development but create long-term maintenance challenges. ATD, in particular when self-admitted, impacts the foundational structure of software, making it difficult to detect and resolve.

This study investigates the lifecycle of ATD, focusing on how it affects i) the connectivity between classes and ii) the frequency of file modifications. We aim to understand how ATD evolves from introduction to repayment and its implications on software architectures.

Our empirical approach was applied to a dataset of SATD items extracted from various software artifacts. We isolated ATD instances, filtered for architectural indicators, and calculated dependencies at different lifecycle stages using FAN-IN and FAN-OUT metrics. Statistical analyses, including the Mann-Whitney U test and Cliff's Delta, were used to assess the significance and effect size of connectivity and dependency changes over time.

We observed that ATD repayment increased class connectivity, with FAN-IN increasing by 57.5\% on average and FAN-OUT by 26.7\%, suggesting a shift toward centralization and increased architectural complexity after repayment. Moreover, ATD files were modified less frequently than Non-ATD files, with changes accumulated in high-dependency portions of the code.

Our study shows that resolving ATD improves software quality in the short-term, but can make the architecture more complex by centralizing dependencies. Also, even if dependency metrics (like FAN-IN and FAN-OUT) can help understand the impact of ATD, they should be combined with other measures to capture other effects of ATD on software maintainability.


\end{abstract}


\section{Introduction}
The technical debt (TD) metaphor in software development refers to compromises made to meet short-term goals that may detract from the long-term health and maintainability of software systems \cite{cunningham1992wycash, avgeriou2016managing}. A specific form of TD is self-admitted technical debt (SATD), where developers consciously make such compromises due to constraints such as time or resource limitations \cite{potdar2014exploratory}. These compromises, often directly expressed in natural language through textual artifacts \cite{sutoyo2024deep}, allow developers to acknowledge areas of the software that require future improvement \cite{da2017using}. Understanding the lifecycle of SATD is crucial for managing its impact on software project health \cite{tan2023lifecycle}. This is particularly the case for architecture technical debt (ATD).



ATD represents trade-offs in the architectural design decisions of software systems, such as technology choices or application of architecture patterns \cite{besker2017impact}. Unlike other types of debt, ATD affects the core structure and behavior of software, making it difficult to detect and resolve without substantial development effort. This unique impact of ATD has been highlighted in several studies. For instance, Besker et al. \cite{besker2016systematic} found that ATD had the highest negative impact on daily software activities, while Lenarduzzi et al. \cite{lenarduzzi2021systematic} underscored the high cost and risk associated with managing ATD compared to other TD types.

Detecting ATD remains a significant challenge due to its complexity and nuanced presence within software architectures. This difficulty is underscored by a recent study by Li et al. \cite{li2023automatic}, which identified only 116 ATD items out of 8,812 technical debt instances across various artifacts—ATD accounted for a mere 0.13\% of the total debt items. Moreover, we independently validated these ATD items and included only those that all authors consistently labeled as true positives. This process revealed that only 57 of the 116 ATD items were correctly identified as true ATD, highlighting the challenges in reliably detecting and classifying architectural debt. These findings emphasize the need for more targeted methods to capture the unique aspects of ATD.

While Tan et al. \cite{tan2023lifecycle} have explored the lifecycle of various types of technical debt by combining data from issue trackers and source code repositories, their broad approach did not address the specific challenges associated with ATD. Although their study provides valuable insights into how technical debt is reported, discussed, and resolved, it overlooks the analysis of specific files within the repository and the dependencies among files associated with particular commits.

Unlike code-level technical debt, which often manifests in clear, tangible issues such as poor naming or code duplication \cite{han2022code}, ATD operates at the high-level design and structure of the software \cite{martini2015danger}. It is inherently more abstract, often embedded in architectural decisions and dependency relationships. Moreover, ATD does not always produce immediate symptoms, such as failing tests or runtime errors. Instead, its impact (such as reduced scalability or maintainability) typically becomes evident only as the system evolves or scales.

Despite prior efforts to analyze TD, the unique challenges of detecting and resolving ATD remain complex \cite{verdecchia2020architectural}. Existing tools often focus on high-level indicators of TD \cite{avgeriou2020overview, sutoyo2024development} without capturing the deeper structural dependencies impacted by ATD. As a result, software engineers face difficulty predicting the long-term impact of ATD on system maintainability and architectural stability. This paper addresses this gap by focusing on the lifecycle of ATD, applying dependency metrics to provide actionable insights into its evolution and impact.

Given the unique challenges posed by ATD, conducting a dependency analysis of specific files within a repository and examining dependencies associated with particular commits can offer essential insights that general TD analysis methods may miss. Software systems are typically composed of interdependent components \cite{claes2018intercomponent}, and the propagation of ATD can lead to complex, cascading effects across these dependencies. By not considering these dependencies, prior studies may overlook significant architectural impacts of ATD, leading to an incomplete understanding of its lifecycle and effects on system quality.


Dependency analysis, particularly class connectivity through metrics like FAN-IN and FAN-OUT, is crucial for understanding the propagation and concentration of ATD within a software architecture \cite{sharma2016designite}. FAN-IN measures the number of modules or files that depend on a specific file, while FAN-OUT tracks how many other modules or files a particular file depends upon \cite{murgia2012refactoring}. High FAN-IN values indicate critical files that serve as central nodes in the software \cite{nasseri2010empirical}, making them potential hotspots for ATD accumulation, as technical debt in these files may cascade to dependent modules. Similarly, high FAN-OUT values suggest files with extensive dependencies \cite{mubarak2010evolutionary}, which could signify complex modules that are difficult to understand and maintain \cite{kemerer1995software} and are more frequently targeted for refactoring than low-coupled classes \cite{murgia2012refactoring}.

The key contributions of this paper are the following:
\begin{enumerate}
    
    \item An investigation into the lifecycle management of ATD through dependency analysis, addressing how ATD items evolve within a software architecture.
    
    
    \item Demonstration of the utility of FAN-IN and FAN-OUT metrics in analyzing the structural evolution of software systems, providing insights into how architectural changes influence connectivity and complexity.

    \item A comprehensive comparison of ATD and Non-ATD items, showcasing differences in dependency structures, modification frequencies, and their correlation with complexity over time.
    
\end{enumerate}

This study is organized as follows: Section~\ref{sectionBG} introduces the background, while Section~\ref{sectionRM} describes the study design. Section~\ref{sec:_worked_example}, ~\ref{sec:result_rq1}, and \ref{sec:result_rq2} present and elaborate on the results for a worked example, RQ1 and RQ2, respectively. Meanwhile, Section~\ref{sectionD} discusses these findings. Finally, Section~\ref{sectionC} provides a conclusion and outlines future work.

\section{Background}
\label{sectionBG}
This section divides our background into two parts: architecture technical debt and FAN-IN and FAN-OUT metrics.
\subsection{Architecture Technical Debt (ATD)}
Alves et al. \cite{alves2014towards} categorized 13 types of TD and their indicators, including architecture technical debt (ATD). They defined ATD as issues arising in the software architecture, identifying structure dependencies/analysis, and violation of modularity as key indicators. Subsequently, Li et al. \cite{li2020identification} refined the debt types and indicators of TD from the previous framework \cite{alves2014towards}. To make this article self-contained, the definition of ATD indicators based on that framework can be described as follows: 

\begin{itemize}
\item Violation of modularity \textit{(VioMod)} occurs when multiple modules become interdependent instead of remaining independent. 

\item Using obsolete technology \textit{(ObsTech)} indicates cases where architecturally-significant technologies have become obsolete.
\end{itemize}


\subsection{FAN-IN and FAN-OUT Metrics}
\textit{VioMod} can lead to increased coupling within the system. One effective way to assess and quantify these violations is through the use of FAN-IN and FAN-OUT metrics, which act as proxies for identifying the degree of coupling between classes. Specifically, FAN-IN measures how many other classes depend on a given class, while FAN-OUT quantifies how many classes a given class depends upon \cite{murgia2012refactoring}.

The impact of coupling on TD is particularly significant in architectural contexts. Studies have shown that high coupling, particularly in core architectural components, can lead to increased defect-related maintenance costs and lower system stability. For instance, tightly coupled core components are often more expensive to maintain than loosely coupled ones, indicating that refactoring efforts should prioritize these areas to reduce debt \cite{maccormack2016technical}. 

In their study, Murgia et al. \cite{murgia2012refactoring} examined the relationship between refactoring practices and class coupling, specifically through FAN-IN and FAN-OUT metrics, across four open-source systems. They revealed that highly coupled classes, especially those with significant FAN-OUT, are more frequently targeted for refactoring than low-coupled classes. Similarly, Imran \cite{imran2019design} investigated parameter-based refactoring and its impact on software coupling, specifically in terms of FAN-IN and FAN-OUT metrics. By focusing on modifications that adjust method parameters, the study highlighted how refactoring can reduce coupling in highly interdependent classes. 

In addition, FAN-IN and FAN-OUT metrics provide deeper insights into program dependencies by capturing runtime interactions, which can more accurately reflect the real coupling levels in complex systems \cite{wang2007dynamic}. Beyond the scope of individual classes, high FAN-IN and FAN-OUT across all modules involved in the execution paths of an activity indicate a broader dependency impact, significantly reducing the evolvability of an activity and its application, thus reflecting a greater magnitude of TD \cite{vora2022measuring}.

\section{Study Design}
\label{sectionRM}

This section details the process of selecting and analyzing ATD from an existing dataset provided by Li et al.~\cite{li2023automatic}. The overall approach is articulated in the following steps: 

\begin{table}[htpb]
    \caption{Dataset from Li et al. \cite{li2023automatic}} 
    \begin{tabular}{p{1.6cm} p{1.4cm} p{1.2cm} p{1.2cm} p{1.2cm}}
    \toprule
    SATD Type & CC & IS & PS & CM\\
    \midrule
    C/D       & 2,703        & 2,169  & 510  & 522    \\
    DOC       & 54           & 487   & 101  & 98     \\
    TEST      & 85           & 338   & 68   & 58     \\
    REQ       & 757          & 97    & 20   & 27     \\
    DEFECT    & 472          & 25    & 1    & -      \\
\rowcolor{lightgray}     ATD       & -            & 93    & 10   & 13     \\
    BUILD     & -            & 67    & 8    & 29     \\
    \midrule
    Non-Debt  & 58,676       & 19,904 & 42,82 & 4,253  \\ 
    \bottomrule
    \end{tabular}
    \label{tab:tb_dataset_li}
\end{table}

\textbf{Dataset overview} -- In their recent study on SATD, Li et al. \cite{li2023automatic} released an extensive dataset aimed at automated SATD detection (see Table~\ref{tab:tb_dataset_li}). The dataset consists of 4 different software artifacts across 103 Apache OSS projects. It includes 5k commit messages \textbf{(CM)} and 5k pull requests section \textbf{(PS)}. It also contains 23k issues section \textbf{(IS)} obtained from their previous study \cite{li2022identifying}. Each instance is categorized by type: code/design (C/D), requirement (REQ), documentation (DOC), test (TEST), defect (DEFECT), build (BUILD), architecture (ATD) debt, and Non-Debt. Finally, the dataset includes 62k code comments \textbf{(CC)} from a dataset by Maldonado et al. \cite{da2017using}, annotated with specific types of SATD, such as design, requirement, documentation, and test debt.




\textbf{Selection of ATD items} -- Our research focuses exclusively on datasets that include ATD (highlighted with the \textcolor{darkgray}{\textbf{gray}} color in Table~\ref{tab:tb_dataset_li}). Thus, only data from commit messages, issue sections, and pull requests are used, as ATD items are unavailable in source code comments.

Table~\ref{tab:tb_dataset} presents the ATD dataset filtering process based on Li et al. \cite{li2023automatic}. Initially, this dataset contained 4,071 CC, 3,276 IS, 718 PS, and 747 CM identified as potential SATD sources. From this dataset, we focused on ATD items, selecting 93 from IS, 10 from PS, and 13 from CM, resulting in a total of 116 ATD items. The numbers in parentheses indicate the initial number of ATD items identified from each artifact.

\begin{table}[htpb]
    \caption{ATD dataset provided by Li et al. \cite{li2023automatic}} 
    \begin{tabular}{p{2.4cm} p{1.2cm} p{1.2cm} p{2.2cm}}
        \toprule
        \multirow{2.5}{9em}{Source of Artifact} & \multicolumn{2}{c}{ATD Indicator} & \multirow{2.5}{9em}{\# of ATD}\\
        \cmidrule{2-3}
         & VioMod & ObsTech & \\
        \midrule
        
        IS & 22 \textit{(42)} & 23 \textit{(51)} & 45 \textit{ (93)} \\
        PS & \ 3 \textit{(10)} & \ 0 \textit{\ \ (0)} & \ 3 \textit{ (10)} \\
        CM & \ 9 \textit{(12)} & \ 0 \textit{\ \ (1)} & \ 9 \textit{ (13)} \\
        \hline
        Total & 34 \textit{(64)} & 23 \textit{\:(52)} & 57 \textit{(116)} \\
        \bottomrule
    \end{tabular}
    \label{tab:tb_dataset}
\end{table}

\textbf{Dataset filtering and re-labeling} -- To improve the accuracy of the dataset and minimize false positives, each author independently validated these ATD items using the Li et al. \cite{li2020identification} classification framework. Instead of using Cohen’s kappa for inter-rater reliability, we included only those items that \textit{all} three authors consistently labeled as true positives. Our study began with an initial set of 116 ATD items, this re-labeling process resulted in a final dataset of 57 ATD items: 45 from IS, 3 from PS, and 9 from CM. 

\textbf{Selection of \textit{Violation of modularity} ATD} -- The dataset comprises two subsets: Violation of modularity (\textit{VioMod}) items and Using obsolete technology (\textit{ObsTech}) items. As the focus of this work focuses solely on the VioMod subset, we included 22 items from IS, 3 from PS, and 9 from CM (see Table~\ref{tab:tb_dataset}). 

For each VioMod item, we attempted to locate the associated commit hash in the repository. However, we could not find the corresponding hashes for 10 items and 2 items from IS and PS, respectively. Finally, this process successfully linked 12 ATD items from IS, 1 from PS, and 9 from CM to their corresponding commits. Since Li et al. \cite{li2023automatic} treated the summary, description, and comments within a single issue as separate debt items, we identified 4 duplicated ATD items originating from the same issue ID. To avoid redundancy, we included only one item from each unique issue ID. In summary, the final dataset used in this paper consists of 8 items from IS, 1 from PS, and 9 from CM. The complete dataset is available in our replication package\footnote{\url{https://doi.org/10.5281/zenodo.14697268}}.

Despite the limited number of identified ATD items, the study focuses on high-quality, architecture-relevant data derived from multiple textual sources, including CM, PS, and IS artifacts. Each ATD item has been carefully selected based on its clear association with architectural concerns, ensuring that the analysis remains directly relevant to the research question. This approach guarantees the validity of the ATD items, as they are representative of architectural issues rather than noise or irrelevant debt, thus maintaining the integrity of the dataset. From the 18 identified ATD items, we identified 5,135 files affected in the introduction phase and 3,553 files in the payment phase, providing a robust basis for analyzing architectural impacts over time.



\textbf{Lifecycle analysis per ATD item} -- We followed a structured approach to analyze each ATD item’s lifecycle, using the following steps:

\begin{enumerate}
    \item \textit{Retrieval and tracing}: We began by gathering ATD items from IS, PS, and CM sources. Each item was traced back to its associated source code through its commit hashcode, enabling us to pinpoint when the ATD instances occurred.

    \item \textit{Identification of `introduction' and `payment' moments}: Using \texttt{git log}, we extracted both the code changes and their corresponding commit messages. To identify the moment of \textit{introduction}, we used \texttt{git blame} to trace the initial addition of ATD-related code in the repository, such as when workarounds were introduced or new components (e.g., functions or classes) were added without proper documentation or unit tests. The moment of \textit{payment} was determined by analyzing commits that addressed the ATD, such as removing workarounds, resolving modularity issues, or updating compatibility with frameworks or libraries \cite{verdecchia2020architectural}. This approach follows the steps outlined by Tan et al. \cite{tan2023lifecycle}, providing a complete view of the lifecycle of each item.

    \item \textit{Historical state verification}: With \texttt{git checkout} followed by a specific hashcode, we navigated to the repository's historical state, reflecting the state of the project at ATD \textit{introduction} and \textit{payment}. This enabled verification of deployment changes over time.

    \item \textit{Calculation of `payment interval'}: After identifying the ATD \textit{payment} moment, we calculated the interval from \textit{introduction} to \textit{payment} by measuring the time between the initial commit date and the final commit date that addressed the ATD issue.

    \item \textit{Dependency analysis with FAN-IN and FAN-OUT metrics}: At both \textit{introduction} and \textit{payment} moments, we calculated the dependency changes using FAN-IN and FAN-OUT, measured with the Understand tool by Scitools \cite{scitools2024}. This analysis helped us observe the impact of ATD on software complexity by tracking dependency structure changes between classes over time \cite{mubarak2010evolutionary}.

    \item \textit{Source code isolation}: To avoid data complexity, we limited our analysis to ATD-affected files, excluding unrelated files from the dataset. Additionally, we focused exclusively on source code files, ommitting non-source files such as HTML, markdown, and text files, as they do not contribute to ATD-specific modifications.
\end{enumerate}

\textbf{Comparison with Non-ATD items} -- To gain a more holistic understanding of the files, we randomly selected 18 Non-ATD samples—referred to as Non-Debt items in Li et al. \cite{li2023automatic}, provided we could locate their corresponding commit hash codes in the repository. We chose 18 items to match the number of ATD items analyzed, and to ensure a balanced comparison. In addition, to differentiate the terms used for Non-ATD items, we substituted the `payment' and `introduction' phases with the \textit{recorded} and \textit{initial commit} phases. 

\subsection{Objectives and Research Questions }
This research aims to investigate the lifecycle of architecture technical debt from various source artifacts and analyze it employing the dependency (FAN-IN and FAN-OUT) approach \cite{scitools2024}. Using the Goal-Question-Metric (GQM) \cite{van2002goal} formulation, the objective is stated as follows:

\begin{displayquote}
\textit{
\textbf{Analyze} commit messages, pull requests, and issue tracking systems \textbf{for the purpose of} investigating the lifecycle of architecture technical debt items \textbf{with respect to} their introduction and payment phases, the number of changes between those phases, and the FAN-IN and FAN-OUT metrics of ATD-affected files \textbf{from the point of view of} software engineers \textbf{in the context of} open-source software.
}
\end{displayquote}

The goal can be further refined into the following two research questions (RQs):\\

\textbf{RQ1}: How does architecture technical debt affect the connectivity of classes within the software system?\\
\underline{\textit{Metrics associated}}: FAN-IN and FAN-OUT of the affected files and a software metric, i.e., source lines of code (SLOC)

\textbf{RQ2}: How does the number of changes in files associated with ATD items correlate with dependencies and complexity?\\
\underline{\textit{Metrics associated}}: Number of changes per file between introduction and payment, and FAN-IN, FAN-OUT, and cyclomatic complexity in the payment phase

\subsection{Tools and Technologies}
The tools and technologies used in the study are carefully chosen to handle different aspects of data extraction, analysis, and statistical evaluation. PyDriller \cite{PyDriller} is employed to mine the \texttt{Git} commit history, such as the author of changes, timestamps, last known path, and the specific files affected. By using PyDriller, the \textit{introduction} and \textit{payment} dates of ATD items, as well as the \textit{initial commit} and \textit{recorded} dates of Non-ATD samples, can be traced accurately, and detailed commit-level data can be collected for further analysis.

To calculate FAN-IN and FAN-OUT values of ATD and Non-ATD for the relevant classes, we utilize Understand by Scitools \cite{scitools2024}, a widely used static analysis tool known for its ability to parse and analyze code to provide detailed metrics. This tool allows for precise calculation of FAN-IN and FAN-OUT, which are key metrics for assessing the connectivity of classes. By obtaining these values at the introduction and payment stages of ATD, the study will be able to assess the impact of technical debt on class dependencies.

Furthermore, Lizard\footnote{\url{https://github.com/terryyin/lizard}} is employed to automatically extract fundamental code metrics, including cyclomatic complexity, which quantifies the logical complexity of code. This tool examines source code to determine the number of independent paths within the program, offering an indication of the code’s complexity and potential challenges in terms of maintenance and comprehensibility.
\section{Results - Worked example}
\label{sectionR}
\label{sec:_worked_example}
In this subsection, we illustrate our approach with a specific example of one ATD item identified in the issue tracking system, shortened as ATD \#1\footnote{Issue ID \texttt{CAMEL-789}, \url{https://issues.apache.org/jira/browse/CAMEL-789}}. This issue affected 33 Java files within the Apache Camel project and was linked to the repository via commit hash \texttt{\#77b260b6}\footnote{\url{https://github.com/apache/camel/commit/77b260b6}}. Table \ref{tb_camel-789-files} provides a subset of six affected Java files, highlighting the key metrics FAN-IN and FAN-OUT, at both the debt \textit{introduction} and \textit{repayment} stages.

Notably, during the repayment phase, three files (\texttt{ChoiceBuilder.java}, \texttt{WhenBuilder.java}, and \texttt{ToBuilder.java}) were removed, potentially reflecting a restructuring effort aimed at reducing complexity and improving maintainability. This change illustrates how debt repayment efforts can also modify the dependency structure, with certain files either removed or their dependencies altered. When files were deleted or their dependencies modified, FAN-IN and FAN-OUT data became unavailable. Excluding those removed files resulted in an \textit{unpaired} dataset, which influenced our analytical approach. Given that the data does not follow a normal distribution, as confirmed by the Anderson-Darling \cite{nelson1998anderson} normality test, we employed the Mann-Whitney U test\footnote{Mann-Whitney U statistical test \cite{mcknight2010mann} is a non-parametric statistical test suited for situations where data may not follow a normal distribution.} to evaluate the differences in FAN-IN and FAN-OUT values between the introduction and repayment phases.

Additionally, to determine practical significance, we employ the effect sizes using Cliff's delta ($\delta$) as a measure to assess the impact on the modular structure and connectivity within the software system. Cliff's $\delta$ \cite{cliff2014ordinal} quantifies the standardized difference between two means, with values interpreted as follows: a negligible effect if \(|\delta| < 0.147\), a small effect if \(0.147 \leq |\delta| < 0.33\), a medium effect if \(0.33 \leq |\delta| < 0.474\), and a large effect if \(|\delta| \geq 0.474\). 

\begin{table}[htpb!]
\caption{Subset of Java Files from ATD \#1}

\begin{tabular}{lp{0.55cm}p{1cm}p{0.55cm}p{1cm}}
\toprule
\multirow{3}{*}{Filename} & \multicolumn{2}{c}{Introduction} & \multicolumn{2}{c}{Payment}\\
\cmidrule{2-5}
& F-IN & F-OUT & F-IN & F-OUT \\
\midrule
.../Endpoint.java & 11 & 0 & 196 & 5 \\
.../RuntimeCamelException.java & 2 & 0 & 66 & 0 \\
.../processor/SendProcessor.java & 2 & 2 & 6 & 9 \\
.../builder/ChoiceBuilder.java & 2 & 4 & N/A & N/A \\
.../builder/WhenBuilder.java & 1 & 6 & N/A & N/A \\
.../builder/ToBuilder.java & 1 & 5 & N/A & N/A \\
\bottomrule
\label{tb_camel-789-files}

\end{tabular}
\end{table}

\begin{figure*}[htb!] 
\centerline{\includegraphics[trim=0.2cm 0.2cm 0.2cm 0.2cm, clip, width=0.68\textwidth]{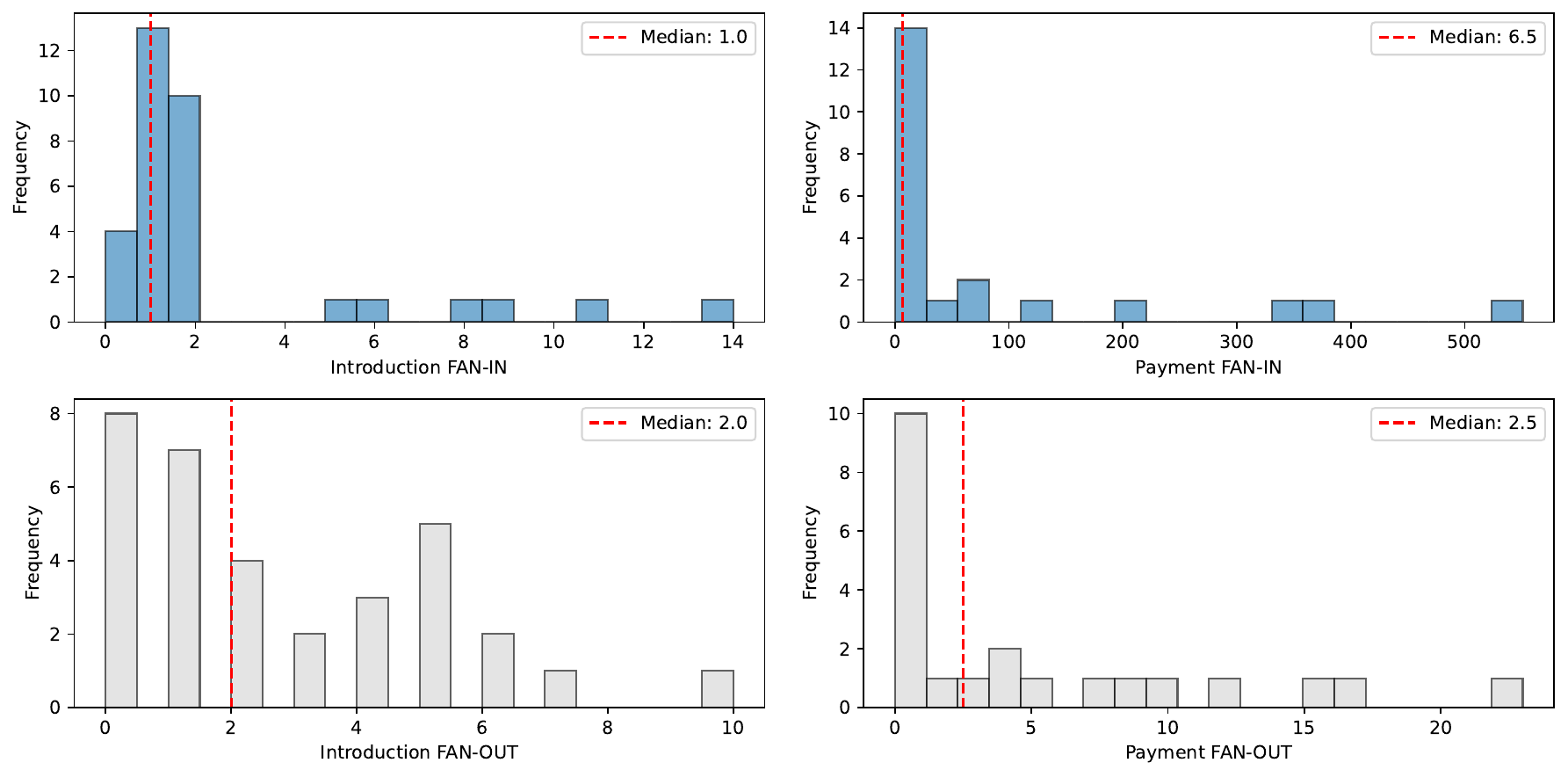}}
\caption{Data distribution of FAN-IN and FAN-OUT for ATD \#1}
\label{fig_fan-in-out-camel-789}
\end{figure*}

Figure \ref{fig_fan-in-out-camel-789} illustrates the distribution of FAN-IN and FAN-OUT values at the ATD introduction and repayment, respectively. The statistical analysis indicates the rejection of the null hypothesis ($H_0$: `\textit{the values of FAN-IN during the ATD introduction and its payment come from the same distribution}') at $\alpha=0.05$. The directional test shows that we cannot reject the $H_1$: `\textit{the values of FAN-IN during repayment are larger than during introduction}', with the median growing from 0.69 at the introduction to 2.01 at payment (\textit{p}-value = $0.0002$). As ATD was resolved, the files affected by this issue experienced an increase in dependency centralization. Cliff's $\delta$ for FAN-IN, calculated at $0.5868$, indicates a large effect size, suggesting that files affected by ATD experience a significant increase in FAN-IN in the payment compared to the introduction phase.

Conversely, we could not reject the null hypothesis for the FAN-OUT ($H_0$: `\textit{the values of FAN-OUT during the ATD introduction and its repayment come from the same distribution}') at $\alpha=0.05$. This smaller shift in FAN-OUT, with Cliff's $\delta$ of $0.1612$ (i.e., a small size), suggests that while there is a slight increase in outgoing dependencies, the impact is less pronounced compared to FAN-IN.

Analyzing ATD \#1 from the Apache Camel project showed that resolving the debt increased incoming dependencies (FAN-IN), making the system more interconnected, while outgoing dependencies (FAN-OUT) changed only moderately. However, it is important to note that this increase in FAN-IN may not be solely attributable to the resolution of ATD; it could also be influenced by the natural evolution of the system over time or other unrelated factors.




\section{Results - RQ1}
\label{sec:result_rq1}




For all the items that were successfully identified as ATD instances from the \textit{VioMod} indicator, we ran a similar analysis to the example shown above. Table \ref{tab:tb_statistic_summary} summarizes the results for all the ATD and Non-ATD items in our dataset. Key observations are outlined below:



\begin{table}[htpb]
\centering
\caption{Summary of FAN-IN and FAN-OUT Metric Changes}
\begin{tabular}{lp{1.2cm}p{1.1cm}p{1.2cm}p{1.1cm}}
\toprule
\multirow{4}{*}{\textbf{Metric}} & \multicolumn{2}{c}{\textbf{ATD Items}} & \multicolumn{2}{c}{\textbf{Non-ATD Items}} \\
\cmidrule{2-5}
& \% Change & Cliff's $\delta$ & \% Change & Cliff's $\delta$ \\
\midrule
Number of files & -44.5\% & N/A & -21.26\% & N/A\\
FAN-IN (Avg.)  & +57.5\% & 0.1028 & +89.06\% & 0.2766 \\
FAN-IN (Max)   & +49.33\% & N/A & +86.32\% & N/A \\
FAN-OUT (Avg.) & +26.7\% & 0.1213 & +36.18\% & 0.1805 \\
FAN-OUT (Max)  & +21.23\% & N/A & +3.87\% & N/A \\
\bottomrule
\end{tabular}
\label{tab:tb_statistic_summary}
\end{table}


\begin{itemize}
    \item \textbf{File reductions:} The number of files decreased for both ATD and Non-ATD items, but the reduction was greater for ATD items (44.5\%) than for Non-ATD items (21.26\%). This may indicate more substantial refactoring or removal of obsolete code in ATD-related efforts.
    \item \textbf{FAN-IN changes:} Non-ATD items exhibited a larger average FAN-IN increase (+89.06\%) compared to ATD items (+57.5\%), with small vs. negligible effect sizes ($0.2766$ vs. $0.1028$). This suggests that Non-ATD items experienced more pronounced increases in incoming dependencies over time.
    \item \textbf{FAN-OUT changes:} Both ATD and Non-ATD items showed modest increases in FAN-OUT metrics, with Non-ATD items exhibiting slightly higher average increases (+36.18\% vs. +26.7\%). Effect sizes for FAN-OUT were negligible for ATD ($0.1213$) and small for Non-ATD ($0.1805$), suggesting minimal impact from ATD resolution but slightly noticeable structural shifts in Non-ATD items.
    \item \textbf{Extreme values:} Maximum FAN-IN values increased similarly for ATD and Non-ATD items, while maximum FAN-OUT changes were minimal for Non-ATD items (+3.87\%) but more notable for ATD items (+21.23\%).
\end{itemize}





\subsection{Statistical Analysis (complete sample)}
We conducted a comparative analysis to investigate the differences in FAN-IN and FAN-OUT metrics between ATD and Non-ATD items during the introduction and payment phases. This comparison aims to understand whether ATD items exhibit distinct patterns in dependency changes compared to Non-ATD items.

When considering all the files contained in the ATD items\footnote{As in Section~\ref{sec:_worked_example}, some FAN-IN and FAN-OUT values could not be recorded during the payment phase due to file deletions or merges. To maintain accuracy, these items were excluded, resulting in unpaired datasets analyzed with the Mann-Whitney U test.} that are part of the complete ATD dataset, we observed \textit{p}-values significantly lower than our standard threshold ($\alpha=0.05$), and for both FAN-IN (\textit{p}-value = \(1.24 \times 10^{-16}\)) and FAN-OUT (\textit{p}-value = \(5.05 \times 10^{-22}\)). Considering the set of all files affected by ATD, we \textit{rejected} the two null hypotheses for FAN-IN and FAN-OUT, indicating in each case a statistically significant difference between the introduction and the payment phases. 

\begin{figure*}[htbp!] 
\centerline{\includegraphics[trim=0.2cm 0.2cm 0.2cm 0.2cm, clip, width=0.68\textwidth]{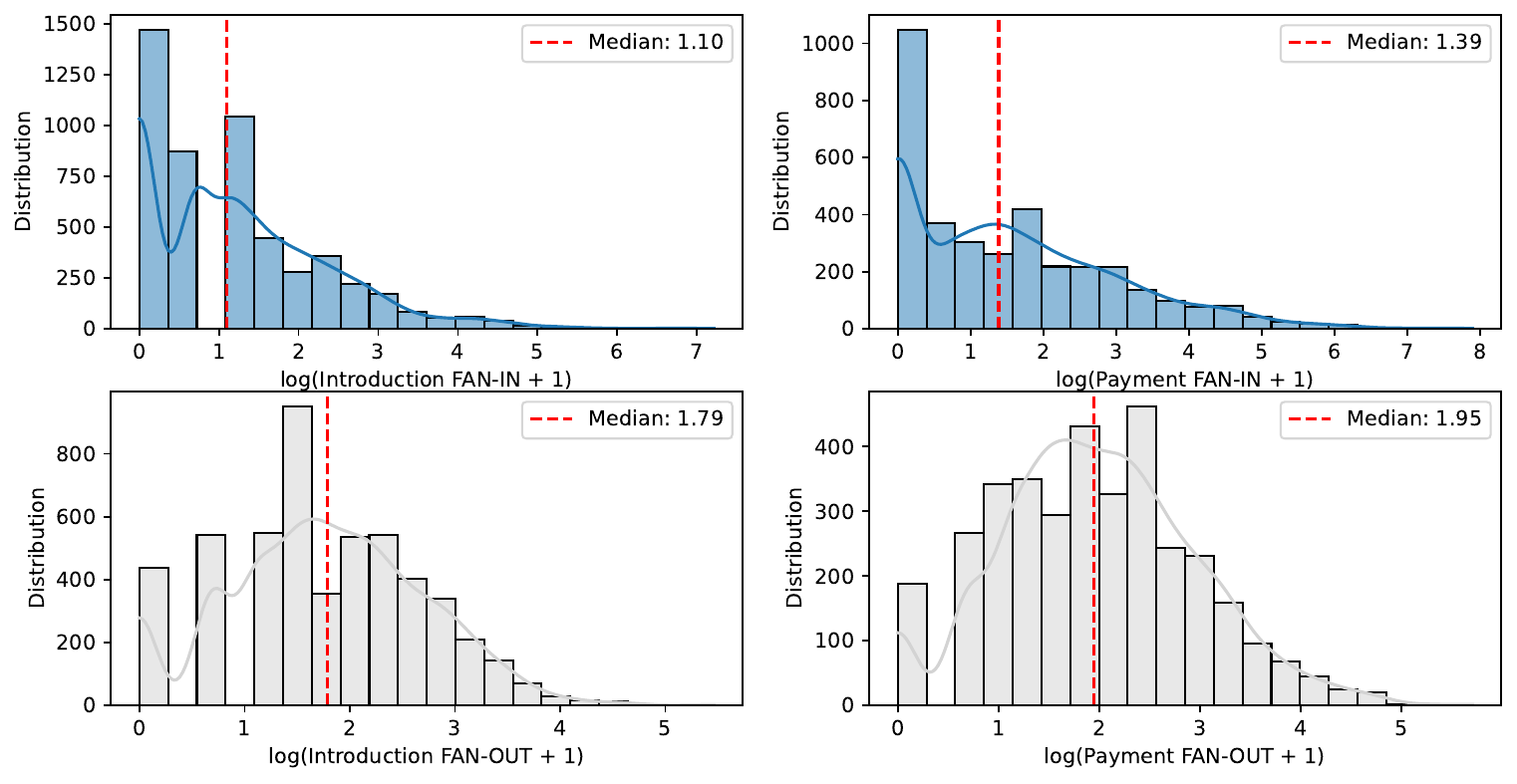}}
\caption{Distribution of FAN-IN and FAN-OUT: Introduction vs. Payment Phases of the ATD items
}
\label{fig_fan-in-out}
\end{figure*}

To understand whether this change is specific to ATD items or part of a general trend in the system, we conducted the same analysis on Non-ATD items. The results also showed statistically significant differences between the initial commit and recorded phases for FAN-IN, \textit{p}-value is \(4.78 \times 10^{-19}\), and for FAN-OUT, \textit{p}-value is \(7.49 \times 10^{-09}\). Since both \textit{p}-values are well below the standard significance level $\alpha$, we rejected the null hypothesis, indicating a significant difference in the FAN-IN and FAN-OUT metrics for Non-ATD items. The results for the ATD items are visualized in Figures~\ref{fig_fan-in-out}\footnote{The figure uses $log1p$ because this transformation effectively handles zero values present in the data, mapping them to zero $(log(1) = 0)$. Without $log1p$, using a standard logarithmic transformation would be problematic, as it is undefined for zero values.}.

For ATD items, the effect size on FAN-IN was evaluated as $\delta = 0.1028$, indicating a negligible effect size. This suggests that the change in FAN-IN between the introduction and payment phases is minimal, meaning that the resolution of ATD items had an insignificant impact on the number of incoming dependencies. Similarly, the effect size on FAN-OUT resulted in $\delta = 0.1213$, which also represents a negligible effect size. While slightly higher than the effect size for FAN-IN, this result still suggests that the overall impact of ATD resolution on outgoing dependencies remains limited.

In comparison, for Non-ATD items, the effect size on FAN-IN was $\delta = 0.2766$, indicating a small effect size. This suggests that the change in FAN-IN between the \textit{recorded} and \textit{initial commit} phases for Non-ATD items is more pronounced than for ATD items. While still relatively small, it implies that Non-ATD items undergo more noticeable structural adjustments in their incoming dependencies. For FAN-OUT, Non-ATD items have an effect size of $\delta = 0.1805$, which falls within the small effect range. This indicates that the change in FAN-OUT between the \textit{recorded} and \textit{initial commit} phases for Non-ATD items, though minor, is more evident than for ATD items, suggesting some small but noticeable shifts in outgoing dependencies.

The comparison between ATD and Non-ATD items highlights that Non-ATD items exhibit a larger effect size for FAN-IN compared to ATD items ($0.2766$ vs. $0.1028$), suggesting that incoming dependencies for Non-ATD items experience greater variation over their lifecycle. Similarly, for FAN-OUT, Non-ATD items show a slightly larger effect size than ATD items ($0.1805$ vs. $0.1213$), indicating that changes in outgoing dependencies are also slightly more noticeable. Overall, these results suggest that ATD resolution has a minimal impact on the structural dependencies of the software system, whereas Non-ATD items show small but more noticeable structural shifts.




\subsection{Statistical Analysis (Individual ATD items)} After evaluating above the overall ATD sample, in this section, we evaluate each ATD item in detail: Table~\ref{tab:tb_mann-whitney_atd} reports the Mann–Whitney U test results for each ATD item, comparing FAN-IN and FAN-OUT between the introduction and repayment phases. Results in parentheses represent the FAN-IN and FAN-OUT metrics normalized by SLOC, which offer a more balanced assessment of connectivity by accounting for differences in code size. For example, we recorded `Y (Y)' for the FAN-IN of the ATD \#1 item to summarize that we could reject the null hypothesis $H_0$ in both cases (e.g., `base' and `normalized' FAN-IN values). For the FAN-OUT of the same ATD item, we recorded `N (N)' as in both cases, we could \textbf{not} reject the null hypothesis. It is important to notice that a statistically significant difference (\textit{p}-value < 0.05, recorded as `Y') suggests that the ATD item likely influenced the connectivity metric, whereas a lack of significance (\textit{p}-value $\geq$ 0.05, recorded as `N') indicates stability in connectivity, despite the presence and resolution of the ATD item.

\begin{table}[htpb!]
\caption{Mann–Whitney U test \textit{p}-value for each ATD Item between Its Payment and Introduction} 
\begin{tabular}{lcc}
\hline
ATD & FAN-IN (FAN-IN/SLOC)                        & FAN-OUT (FAN-OUT/SLOC)                        \\
\midrule
\#1 & Y (Y) & N  (N) \\
\#2  & N  (N)     & N  (N)\\
\#3  & $-$  $(-)$    & $-$  $(-)$\\
\#4  & N  (N)    & N  (N)\\
\#5  & N  (N)    & N  (N)\\
\#6  & $-$  $(-)$    & $-$  $(-)$\\
\#7  & N  (N)    & N  (N)\\
\#8  & N  (N)    & N  (N)\\
\#9  & N  (N)     & N  (N)\\
\#10  & Y (Y)    & N  (N)\\
\#11  & Y (N)  & N (Y)\\
\#12  & N  (N)    & N  (N)\\
\#13  & N  (N)    & N  (N)\\
\#14  & N  (N)    & N  (N)\\
\#15  & N  (N) & N  (N)\\
\#16  & Y (N)    & Y (N)\\
\#17 & Y (N) & Y (N)\\
\#18 & N  (N) & N  (N)\\
\bottomrule 
\multicolumn{3}{c}{(Y for \textit{p}-value < 0.05; N for \textit{p}-value > 0.05)}
\label{tab:tb_mann-whitney_atd}
\end{tabular}
\end{table}

For FAN-IN, the statistical tests for most items resulted in \textit{p}-values $\geq$ 0.05, meaning that we could not reject the null hypothesis for these items. This implies that no statistically significant difference was observed in incoming connections (relative to SLOC) between the introduction and payment phases, suggesting stability in the FAN-IN metric for these ATD items. However, when the \textit{p}-values were < 0.05, allowing us to reject the null hypothesis, the resolution of these specific ATD items (for example items \#1 and \#10) affected the incoming connectivity of the code, potentially indicating a structural or architectural shift. In all the cases where the null-hypothesis was rejected, we detected a significative direction in the relationship: the repayment values are typically higher than the introduction values.

For ATD items \#11, the FAN-IN metric differs between base and normalized values: the base metric shows a significant difference (Y), indicating that absolute incoming connections changed between introduction and payment phases. When normalized by SLOC (FAN-IN/SLOC), there is no significant difference (N), suggesting proportional changes to code size, keeping connectivity density stable. In contrast, FAN-OUT results show no significant difference in base terms (N), but a significant difference (Y) when normalized,  indicating total outgoing connections were stable while connectivity density per line of code was affected by ATD resolution.

ATD items \#16 and \#17 also display differences between base and normalized metrics for both FAN-IN and FAN-OUT. The base FAN-IN and FAN-OUT metrics display significant differences (denoted by Y), while the normalized metrics show no significant differences (denoted by N). This implies that the observed connectivity changes were proportional to the code size, and thus, when normalized, they no longer appear statistically significant.

For ATD items \#3 and \#6, the symbol `$\text{--}$' indicates a unique case where only folders, not specific files, were referenced. As a result, we could not calculate file-specific metrics like FAN-IN, FAN-OUT, or complexity, leading to their exclusion from parts of the analysis.

Meanwhile, Table~\ref{tab:tb_mann-whitney_non_atd} presents the results of the Mann–Whitney U test for each Non-ATD item, comparing FAN-IN and FAN-OUT values between the \textit{recorded} and \textit{initial commit} phases. Like the ATD items, this analysis assesses whether there are statistically significant changes in connectivity metrics, both in absolute terms and normalized by SLOC. For most Non-ATD items, the results show no significant differences in either FAN-IN or FAN-OUT, both in base and normalized metrics. However, there are notable exceptions for Non-ATD items \#14 and \#15 that show a significant difference in base FAN-IN metric (denoted by Y), indicating that this item affected both incoming connections. Interestingly, when normalized by SLOC, these differences remain significant, suggesting a substantial change in connectivity density as well as absolute connectivity.

Additionally, Non-ATD items \#14 and \#15 display a significant difference in the base FAN-OUT metric (denoted by Y), while the normalized FAN-OUT/SLOC metric shows no significant difference (denoted by N). This suggests that although the absolute outgoing connections changed, these changes were proportional to the size of the code, leaving the connectivity density unaffected after normalization.

We conducted Mann-Whitney U tests to evaluate the significance of the observed changes: both ATD and Non-ATD items showed statistically significant differences in FAN-IN and FAN-OUT metrics between the phases. However, effect size analysis revealed that these differences were generally small for both categories, with Non-ATD items having slightly larger effect sizes for FAN-IN. This suggests that while connectivity changes occur across the system, the structural impact of ATD resolution is modest.


\begin{table}[htpb!]
\caption{Mann–Whitney U test \textit{p}-value for each Non-ATD Item between It's Recorded vs Initial commit}
\begin{tabular}{lll}
\hline
Non-ATD & FAN-IN (FAN-IN/SLOC) & FAN-OUT (FAN-OUT/SLOC)\\
\midrule
\#1 & N (N) & N (N)\\
\#2  & N (N)    &  N (N)   \\
\#3  & N (N)    & N (N)\\
\#4  & N  (N)   & N (N)\\
\#5  & N  (N)   & N (N)\\
\#6  & N (N)    & N (N)\\
\#7  & N (N)    & N (N)\\
\#8  & N (N)    & N (N)\\
\#9  &  N (N)   & N (N)\\
\#10  &  N (N)   & N (N)\\
\#11  & N (N)   & N (N)\\
\#12  &  N (N)   & N (N)\\
\#13  & N (N)    & N (N)\\
\#14  & Y (Y)     & Y (N)\\
\#15  &  Y (Y)   & Y (N)\\
\#16  &  N (N)   & N (N)\\
\#17 & N (N) & N (N)\\
\#18 & N (N) & N (N)\\
\bottomrule
\multicolumn{3}{c}{(Y for \textit{p}-value < 0.05; N for \textit{p}-value > 0.05)}
\label{tab:tb_mann-whitney_non_atd}
\end{tabular}
\end{table}



\begin{tcolorbox}[colback=gray!5!white, colframe=black, title=Answer to RQ1, boxrule=0.1pt, fonttitle=\small, fontupper=\small]
ATD repayment raises class connectivity: FAN-IN increased by 57.5\% and FAN-OUT by 26.7\%, centralizing dependencies and mildly increasing architectural complexity. In comparison, Non-ATD items exhibit a more significant average FAN-IN increase of 89.06\% and a FAN-OUT increase of 36.18\%.

\end{tcolorbox}

\section{Results - RQ2}
\label{sec:result_rq2}

\begin{figure}[htbp!] 
\centerline{\includegraphics[trim=0.2cm 0.2cm 0.2cm 0.2cm, clip, width=0.48\textwidth]{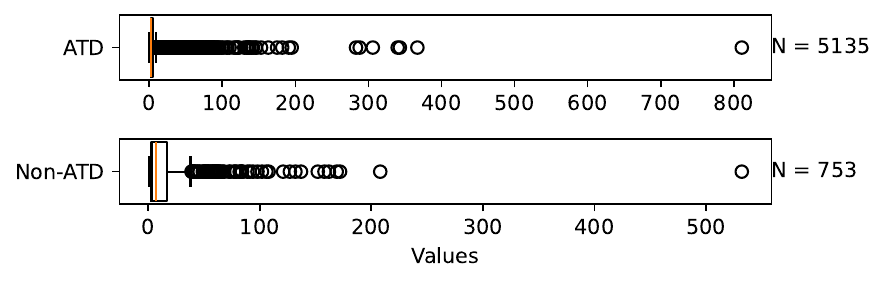}}
\caption{Boxplot of Number of Changes: ATD vs. Non-ATD Files}
\label{fig_number-of-changes}
\end{figure}

In order to answer RQ2, we analyzed the frequency of modifications in ATD-related files compared to Non-ATD files. Our study focused on 18 ATD items, tracking the dates of their initial introduction and final repayment using the \texttt{git blame} tool. We then counted modifications within these timeframes to assess code volatility, as illustrated in Figure~\ref{fig_number-of-changes}.

The analysis of changes in ATD-related files versus Non-ATD files reveals a clear distinction in modification frequency. As shown in Table~\ref{tab:tb_statistic_changes}, our results indicated that ATD-related files underwent significantly fewer frequent modifications, with an average of 6.981 changes per file ($SD = 20.935$), compared with Non-ATD with an average of 16.115 changes per file ($SD = 30.320$).

\begin{table}[htpb]
\centering
\caption{Changes Statistic of ATD vs Non-ATD Items}
\begin{tabular}{llll}
\toprule
Metric & ATD & Non-ATD \\
\midrule
Avg. change  & 6.981  & 16.115 \\
Median change & 2.0   & 7.0 \\
Min. change & 0 & 1 \\
Max. change & 811 & 532 \\ 
Std. deviation & 20.935 & 30.320 \\ 
\hline
\end{tabular}
\label{tab:tb_statistic_changes}
\end{table}

\subsection{Correlation between Dependencies, Complexity, and Number of Changes}
To investigate whether the number of changes plays a role in the increase or decrease of connectivity metrics (FAN-IN and FAN-OUT) and complexity, we measure Partial Spearman's Correlation $r$ \cite{ref1} between FAN-IN and FAN-OUT, cyclomatic complexity, and the number of changes, while controlling for source lines of code (SLOC) for both ATD and Non-ATD items. Controlling for SLOC is essential because larger files typically exhibit higher complexity and connectivity metrics simply due to their size. Without accounting for SLOC, it becomes challenging to determine whether observed changes in FAN-IN, FAN-OUT, or cyclomatic complexity are due to genuine structural or architectural modifications or are merely artifacts of an increase in file size. 

By measuring partial correlations rather than simple Spearman correlations, we account for SLOC’s influence on connectivity, complexity, and change metrics. By controlling for SLOC, we can isolate the relationships among connectivity, complexity, and the number of changes, ensuring that any observed correlations more accurately reflect structural dynamics rather than variations in code size alone.



\begin{table}[h]
\centering
\caption{Partial Spearman's Correlation between FAN-IN, FAN-OUT, Cyclomatic Complexity and Number of Changes while Controlling SLOC}
\begin{tabular}{llll}
\toprule
Metric & Coefficient $r$ & CI95\% & \textit{p}-value\\
\midrule
\multicolumn{4}{c}{ATD}\\
\midrule
FAN-IN & 0.241 & [0.21, 0.27] & <0.001\\
FAN-OUT & 0.175 & [0.14, 0.21] & <0.001\\ 
Complexity & -0.089 & [-0.12, -0.06] & <0.001\\

\midrule
\multicolumn{4}{c}{Non-ATD}\\
\midrule
FAN-IN & 0.298 & [0.22, 0.37] & <0.001\\
FAN-OUT & 0.533 & [0.47, 0.59] & <0.001\\ 
Complexity & 0.313 & [0.24, 0.38] &  <0.001\\
\hline
\end{tabular}
\label{tab:tb_partial_spearman}
\end{table}

Table~\ref{tab:tb_partial_spearman} presents Partial Spearman's correlation coefficients ($r$) between connectivity metrics (FAN-IN and FAN-OUT), cyclomatic complexity, and the number of changes, while controlling for SLOC for both ATD and Non-ATD items. This analysis provides insights into the relationships between structural metrics and change frequency in code, highlighting differences in correlation patterns of ATD and Non-ATD items.

For ATD items, there is a positive and statistically significant correlation between FAN-IN and the number of changes ($r = 0.241$, Confidence Interval (CI95\%) = [0.21, 0.27], $p < 0.001$). This indicates that as incoming dependencies (FAN-IN) increase, the frequency of changes to the code also tends to increase. FAN-OUT also shows a positive correlation with the number of changes ($r = 0.175$, CI95\% = [0.14, 0.21], $p < 0.001$), though the strength of this relationship is slightly weaker than for FAN-IN. Cyclomatic complexity, on the other hand, exhibits a negative correlation with the number of changes ($r = -0.089$, CI95\% = [-0.12, -0.06], $p < 0.001$), suggesting that more complex ATD items may be less frequently modified, possibly due to the challenges associated with modifying highly complex code \cite{mantyla2008types, lopes2022and}.

In contrast, Non-ATD items show a stronger positive correlation between FAN-IN and the number of changes ($r = 0.298$, CI95\% = [0.22, 0.37], $p < 0.001$), indicating that incoming dependencies have a greater impact on change frequency in Non-ATD items compared to ATD items. FAN-OUT has an even stronger correlation with the number of changes ($r = 0.533$, CI95\% = [0.47, 0.59], $p < 0.001$), suggesting that outgoing dependencies in Non-ATD items are closely associated with higher modification frequency. Additionally, cyclomatic complexity for Non-ATD items is positively correlated with the number of changes ($r = 0.313$, CI95\% = [0.24, 0.38], $p < 0.001$), indicating that Non-ATD code is modified more frequently.

\begin{tcolorbox}[colback=gray!5!white, colframe=black, title=Answer to RQ2, boxrule=0.1pt, fonttitle=\small, fontupper=\small]
ATD-related files experience fewer changes compared to Non-ATD files and show slightly weaker correlations with FAN-IN and FAN-OUT. Cyclomatic complexity is negatively correlated with changes in ATD files but positively correlated in Non-ATD files.
\end{tcolorbox}





\section{Discussion}
\label{sectionD}



In this section, we discuss the implications of our findings for practitioners and researchers. Additionally, we address some key aspects of managing ATD: first, the limitations of natural language processing (NLP) methods in capturing structural ATD; second, how addressing ATD can enhance quality metrics but also increase architectural complexity by centralizing dependencies. Third, the limitations of FAN-IN and FAN-OUT at assessing the impact of ATD.

\subsection{Implications for practitioners}
Practitioners managing ATD should balance immediate quality improvements with the potential for increased dependencies that could impact future maintainability. Our results show that while resolving ATD often enhances software quality in the short term, it can also centralize dependencies, creating more tightly coupled architectures. This shift introduces layers of complexity, which may lead to long-term maintenance challenges. Practitioners should therefore adopt a holistic approach that considers both immediate benefits and potential future implications.

Dependency analysis offers a practical approach for understanding and monitoring ATD’s impact on system architecture. Metrics such as FAN-IN and FAN-OUT can help identify \enquote{hotspots} where ATD resolution has led to increased complexity, indicating areas that may require continuous attention even after the debt is repaid. By analyzing these dependency concentrations, practitioners can ensure that short-term quality improvements do not inadvertently compromise modularity and system scalability.

Tracking ATD is inherently complex, and practitioners should consider structured approaches, ideally leveraging automated tools to improve the accuracy of ATD identification and prioritization. Such automation facilitates more effective ATD management by providing a comprehensive view of architectural dependencies and complexities over time.

\subsection{Implications for researchers}
Our research used a dataset filtered for high-confidence ATD instances, yet the sample size remained limited. Future studies could expand on this work by incorporating a larger and more diverse set of open-source and industry-specific projects. A broader dataset would allow for cross-comparison across different software architectures, project scales, and development environments. This could reveal more nuanced patterns in ATD propagation, resolution, and its impact on system architecture.

Our use of FAN-IN and FAN-OUT metrics provided valuable insights into ATD's effect on class connectivity. However, these metrics alone may not fully capture the complexity of ATD's influence. Researchers could explore additional metrics, such as architectural smells, coupling, cohesion, or other graph-based measures, to develop a more comprehensive view of ATD's impact on software architecture. Enhanced metrics might offer a more detailed characterization of ATD-related changes and assist in developing predictive models for ATD accumulation and its structural implications.

\subsection{Key aspects of managing ATD}
\paragraph*{Challenges in ATD detection} Accurately identifying ATD remains a complex task. Although NLP techniques have proven effective in detecting SATD through textual indicators in code comments, commit messages, and issue sections \cite{codabux2021technical, li2023automatic, sutoyo2024deep}, they may not fully capture the intricacies of architectural technical debt. These methods often struggle to capture the more abstract, structural nature of ATD, which is not always explicitly documented or self-admitted. Architectural debt may emerge subtly, through dependencies, code organization, and design decisions that NLP alone may not detect. This challenge suggests the need for an integrated approach that combines NLP with architectural analysis techniques to better identify and track ATD within evolving systems.

Although ATD represents only a small fraction of technical debt, this study shows the challenges of analyzing high-confidence ATD items in real-world projects, which underscores the value of the dataset. With 18 ATD and 18 Non-ATD items, the research is a solid foundation for further exploration. As a proof of concept, it demonstrates the relationship between ATD and class connectivity using FAN-IN and FAN-OUT metrics, offering meaningful insights into ATD's lifecycle and its impact on software architecture.
    
\paragraph*{ATD resolution and complexity trade-offs} Resolving ATD is generally aimed at enhancing software quality by addressing accumulated issues that degrade the architecture over time. However, our findings indicate that while resolving ATD can improve immediate quality metrics, it may also inadvertently centralize dependencies within the system. This centralization can lead to increased architectural complexity as dependencies converge around fewer core components. As a result, there is a \textit{trade-off} between the immediate benefits of quality improvement and the potential long-term consequences of connectivity and dependency issues. These changes, although addressing current ATD, might introduce new structural challenges that affect future maintainability and the system’s modularity. 

\paragraph*{Limitations of dependency metrics} While FAN-IN and FAN-OUT metrics provide valuable insights into the connectivity and dependency structure affected by ATD, they do not fully capture the scope of architectural complexity. These metrics measure the inflow and outflow of dependencies within classes or modules, offering a useful perspective on modularity and cohesion changes (which are central in the \textit{Violation of Modularity} ATD). However, they lack the depth needed to reveal more intricate architectural dependencies, such as indirect couplings or nuanced inter-module relationships that could affect maintainability. FAN-IN and FAN-OUT metrics can therefore serve as preliminary indicators of ATD’s impact but should be complemented with additional architectural metrics or qualitative assessments for a more complete analysis.

\subsection{Threats to validity}
\paragraph*{Internal validity} One of such threats is related to the tool used for extracting dependencies between files (Understand by Scitools). However, the results may vary if a different static analysis tool was used, as each tool may interpret and export dependencies in slightly different ways. Investigating and analyzing these differences among various dependency analysis tools is planned as part of our future work.

\paragraph*{External validity} The study relies on a limited dataset of 18 ATD and 18 Non-ATD items. While this ensures a detailed analysis, and it is the complete set of ATD items from~\cite{li2020identification}, it may not capture patterns present in larger or more diverse datasets from different types of software projects. To mitigate this limitation, we published both the dataset and the methods in a replication package\footnote{\url{https://doi.org/10.5281/zenodo.14697268}}, encouraging other researchers to validate and extend our findings using their own datasets.

\paragraph*{Construct validity} Our study relies on previously labeled ATD items from textual artifacts, which may include misclassifications due to subjective judgment. The dependency analysis also assumes that increases in FAN-IN and FAN-OUT metrics accurately reflect the presence of ATD, which may not account for all complexity factors in software architecture. To reduce the risk of misclassification, we independently validated a subset of ATD items and cross-verified classifications using multiple reviewers. Additionally, we used metrics such as cyclomatic complexity to complement FAN-IN and FAN-OUT, providing a broader perspective on architectural dependencies.
\section{Conclusion}
\label{sectionC}


This study investigated the detection, lifecycle, and impact of ATD, with particular attention to the evolution of dependency structures. By analyzing ATD instances from their introduction to repayment, our findings reveal that addressing ATD often increases connectivity, with resolved ATD items showing higher levels of incoming dependencies (FAN-IN) and, to a lesser degree, outgoing dependencies (FAN-OUT). This trend suggests that, while ATD repayment may improve immediate code quality and modularity, it can also lead to the centralization of dependencies, increasing architectural complexity and introducing potential future maintenance challenges. Despite these shifts, the effect sizes were negligible, indicating that ATD resolution may not have a substantial impact on the modular structure and connectivity between files in the system.

This research shows the need for a more nuanced approach to ATD management: differently from simpler forms of technical debt, ATD affects core structural elements, making it difficult to detect, assess, and mitigate without impacting broader architectural aspects. Effective ATD management must account for dependency patterns and long-term implications for maintainability: dependency metrics, such as FAN-IN and FAN-OUT, provide valuable insights but should ideally be complemented by additional architectural measures to capture the full scope of ATD's impact on system stability.

Future work should validate these findings across a broader dataset and potentially incorporate additional architectural metrics to better understand ATD’s impact on system structure. While this study analyzed 18 ATD and 18 Non-ATD items, the methodology developed here can be extended to larger datasets, providing a foundation for further exploration.

This research serves as a proof of concept, highlighting the relationship between ATD and class connectivity through FAN-IN and FAN-OUT metrics. Expanding the sample size in future studies could reveal whether the observed patterns hold true across diverse systems and a larger set of ATD items. Additionally, extending this work could offer insights into ATD’s specific effects on various types of software architecture, further refining best practices for technical debt management in complex systems.

\balance
\bibliography{main}
\bibliographystyle{IEEEtran} 
\end{document}